\documentstyle[twocolumn,aps]{revtex}

\newcommand{\beq}{\begin{equation}}
\newcommand{\eeq}{\end{equation}}
\newcommand{\beqa}{\begin{eqnarray}}
\newcommand{\eeqa}{\end{eqnarray}}
\newcommand{\beqar}{\begin{eqnarray*}}
\newcommand{\eeqar}{\end{eqnarray*}}

\def \la {\langle}
\def \ra {\rangle}

\begin{document}

\input epsf

\title{\bf\Large
   Quantum Communication Protocol Employing Weak Measurements
}

\author{
Alonso Botero$^{1, \ast}$ \,
 and \,  Benni Reznik$^{2}$
{\ } \\ $^1$ {\em \small Center for Particle Physics, University
of Texas,
 Austin, Texas 78712} \\
$^2$ {\em \small  School of Physics and Astronomy, Tel Aviv
University, Tel Aviv 69978, Israel} }

\maketitle

\begin{abstract}
{
We propose  a communication protocol exploiting correlations
between two events with a definite time-ordering: a) the outcome
of a {\em weak measurement} on a spin, and b) the outcome of a
subsequent ordinary measurement on the spin. In
our protocol, Alice, first generates a ``code'' by performing weak
measurements on a sample of $N$ spins.
 The sample is sent to  Bob, who later performs a
post-selection  by measuring the spin along either of two certain
directions. The results of the post-selection define the ``key'',
which he then broadcasts publicly. Using both her previously
generated code and this key, Alice is able to infer the {\em
direction} chosen by Bob in the post-selection.  Alternatively, if
Alice broadcasts publicly her code, Bob is able to infer from the
code and the key the direction chosen by Alice for her weak
measurement.
Two possible experimental realizations of the
protocols are briefly mentioned. }
\end{abstract}


Weak measurements~\cite{curious,spin100,noise} are a special class
of quantum measurements explored in recent years by Aharonov,
Vaidman, and others. In one such measurement of any given
observable $A$,
 the disturbance caused to the system is minimized
at the expense of precision in a single trial. Nevertheless, after
a large number trials one can determine statistical averages such
as the expectation value of $A$. The distinctive feature of  weak
measurements has to do with the  observed  averages when the
measured system is {\em  post-selected}. Such averages, the
so-called {\em weak values}, may lie outside the bounds of the
spectrum of $A$~\cite{spin100}. Moreover, since they depend on the
chosen pre- and post-selected ensemble, weak values carry
non-trivial information about the {\em choice} of measurement used
for post-selection. In accordance with causality, these unusual
regularities must therefore be {\em a priori} undetectable, i.e.,
 ``hidden in the noise''~\cite{noise}.
Hence, they can only be extracted {\em a posteriori},   from the
correlations between readings of the measurement and the result of
the post-selection.

In the two protocols suggested in this Letter, such correlations
are used for signaling.
In the first protocol, the receiver, Alice, starts by generating a
{\em code} corresponding to the outcomes of a series of weak
measurements on a large sample of spins. She then hands them to
Bob, who is about to depart on a long voyage, together with an
important question that he can only answer at some later time.
When he is ready to respond, Bob performs an ordinary measurement
(post-selection) on these spins. His response corresponds to the
choice of spin component he then measures; for instance, a
measurement of $\sigma_y$ to signal a ``yes'' (``I will marry
you''), or a measurement of $\sigma_z$ to signal a ``no''. The
random sequence of results obtained in either of these
measurements plays the role of a {\em key}, which he then
broadcasts publicly. This is just enough information for Alice to
bin her previous readings and extract the message from the weak
values. The random key is useless to the ever-jealous Eve; it can
be shown that an optimal  sample size exists which is still too
small for Eve to infer Bob's choice from the statistics of the
key.

The security of this protocol is however compromised if the spins
fall in the hands of the eavesdropper.  This drawback is overcome
in the second protocol, in which instead it is Alice who {\em
sends} the message; for this, she performs on the $N$ spins either
of two possible weak measurements and generates a code
corresponding to her outcomes. She then sends the spins {\em and}
the code to Bob. To decode the message Bob post-selects the spins
by performing a sequence of strong measurements. As in the first
case, the resulting key allows him to bin the code and deduce from
the weak values Alice's choice of measurement. In this case, Eve
cannot intercept the message from the spins without Bob being
aware of her actions.

We begin by analyzing the weak measurement scheme. Suppose that
Alice prepares a sample of $N$ spin-$1/2$ particles,
 in the eigenstate $|x+ \ra$ of $\sigma_x$, and she wishes to perform  a weak
measurement of the spin observable
\begin{equation}
A \equiv  {1\over\sqrt2}(\sigma_x+\sigma_y)\, ,
\end{equation} with
eigenstates $|a\pm\ra$. We let $p$ be the  pointer variable of the
measuring device and $|\phi_i\rangle$ its initial state, with  a
Gaussian wave function $\phi_i(p) =  (2\pi \Delta
p^2)^{-1/4}e^{-p^2/4 \Delta p^2 }$ of
 uncertainty $\Delta p$.
The weak measurement may then be described by the usual
transformation taking an initial product state $ |\Psi_i \ra
\equiv |x+\rangle|\phi_i\rangle$ of the spin and the device, to a
final entangled state: \begin{equation} |\Psi_i\ra \rightarrow
|\Psi_f \rangle = c_+|a+\ra |\phi+\ra + c_{-} | a- \ra |\phi -
\ra\, ,
\end{equation} where $c_{\pm} = \la a\pm | x+\ra$, and $|\phi \pm
\ra$ is $|\phi_i \ra$ shifted in $p$ by $\pm 1$. In contrast
however to an ordinary measurement,  $\Delta p$ is here assumed to
be so large that the shifted states $|\phi \pm \ra$ overlap
considerably. Nevertheless, the expectation value of $p$ is still
shifted by the usual expectation value $\la A\ra $:
\begin{equation} \la \Psi_f| p |\Psi_f \ra = \la x\!+ | A | x+
\ra= {1\over \sqrt2}\, . \end{equation} The idea of a  weak
measurement is thus to extract such systematic shifts of the means
from a large sample of identically prepared spins, ensuring at the
same time a minimal disturbance of any individual spin. This
disturbance is naturally related to the overlap between
 $|\phi+\ra$ and $|\phi-\ra$.

A convenient measure of  disturbance $D$ is given in terms of the
Fidelity ${\cal F}$~\cite{fidelity} as
$
D \equiv 1 - {\cal F} $, where ${\cal F}$ is defined as the
probability of obtaining back the initial state after the weak
measurement, ${\cal F} = \left|\left| \la x+ |\Psi_f \ra \right|
\right|^2$; hence, $D$ may be interpreted as  the probability of
``flipping'' the initial direction of the spin. As one verifies,
$D$ involves the overlap  factor  $\la \phi+|\phi- \ra =e^{-1/2
\Delta p^2 }$: \begin{equation} D = {\Delta A^2 \over 2} ( 1 -
e^{-1 / 2 \Delta p^2 } ) \simeq {1\over 8 \Delta p^2 }\, .
\label{D} \end{equation} We assume in the approximation that  $D$
is  small and take $\Delta A^2 = \la A^2\ra - \la A \ra^2=1/2$.

Together then  with the weakness condition  $D \ll 1$, the
resulting distribution for the pointer variable must be
sufficiently broad that the spectrum of $A$ cannot be resolved.
The distribution  therefore takes essentially a Gaussian form,
centered at the expectation value $\la A \ra$, with a large
uncertainty $\sqrt {\Delta p^2 + \Delta A^2 } \simeq \Delta p$ of
order  $D^{-1/2}$. Since in $N$ identical and independent trials
the expected sample mean is   $\bar{p} = \la A \ra \pm \Delta
p/\sqrt{N}$, Alice can determine   $\la A \ra$ if her sample is
large enough.

But now suppose that after completing her measurements Alice gives
the spins to Bob, who  later performs a second,  now ordinary
measurement on each spin, along some arbitrary direction $\hat{b}$
with possible outcomes $|b\pm\ra$. Each outcome in this {\em
post-selection} is correlated with a  particular response of the
measuring device. To keep track of these correlations it is
convenient to re-express the combined state of the system and
apparatus in terms of the basis $|b\pm \ra$
\begin{equation}
|\Psi_f\ra = |b+\ra |\phi(b+)\ra + |b-\ra |\phi(b-)\ra \, ,
\label{psipost}
\end{equation}
where the conditional response of the apparatus is described by
the unnormalized states $|\phi(b\pm)\ra = \la b\pm | \Psi_f\ra$.
For a given result $|b\ra$,
\begin{equation}
 |\phi(b)\ra = C\left[ {1 + A_w(b) \over 2} |\phi+\ra + {1 -
A_w(b) \over 2} |\phi-\ra \right ]\, ,
\end{equation}
 here $C=\la
b|x+\ra$ and $A_w$  is the {\em weak value of A}, defined  as
\begin{equation}
A_w(b)\equiv {\langle b|A|x+ \rangle \over \la b|
x+ \ra} \, .
\end{equation}
The weak value $A_w(b)$ is the defining property of weak
measurements: as before, when $D$ is small the pointer
distribution is essentially  Gaussian of uncertainty $\simeq
\Delta p$; however, as one can easily show, the location of the
mean $\la p \ra = \frac{\la \phi(b)|\, p\,  | \phi(b)\ra}{\la
\phi(b)|\phi(b)\ra}$ is now given by
\begin{equation}
\la p \ra = { {\rm Re} {A_w}(b) \over  1+ {\delta P(b)\over P(b)}
\, } = {\rm Re} A_w + O(D)
\label{weakval}
\end{equation}
where, in terms of $D$,
\begin{equation}
 {\delta P(b)\over P(b)} =
-{1-|A_w(b)|^2 \over \Delta A^2} D\, . \label{freq}
\end{equation}
The correction ${\delta P(b)\over P(b)}$ is the relative deviation
of the probability $P'(b) = \la \phi(b) |\phi(b) \ra $ of
obtaining $|b\ra$ in the presence of the weak measurement, as
measured from the unperturbed probability $P(b)= |\langle b|\psi
\rangle|^2$. We see therefore that the average conditional reading
becomes the real part of the weak value $ {A_w}(b)$ as we approach
the ideally weak situation in which the  transition probability is
left unaltered by the measurement.

To summarize then: if Alice starts with a large sample of $N$
identical spins, she obtains a broad distribution of pointer
readings with a sample mean $\la A \ra \pm \Delta p/\sqrt{N}$;
however,  were she to know-- {\em  for every single spin}--which
of the two possible outcomes $|b+ \ra$ and $|b-\ra$ was obtained
in Bob's measurement, then she could divide her readings into two
categories, corresponding to   two post-selected sub-samples of
size $N_{\pm} \simeq N P( b\pm)$. Her original  distribution would
then break up as a mixture of  two conditional distributions, with
sample means $A_w(b+)$ and $A_w(b-)$ within errors of $\Delta
p/\sqrt{N_\pm}$.  The break-up is captured by a simple sum
rule~\cite{directions} in the limit $D \rightarrow 0$
\begin{equation}
\la A \ra = P(b)\, {\rm Re} A_w(b) + P(b-)\,{\rm Re} A_w(b-) \, ,
\label{sumrule}
\end{equation}which serves as the basis for our protocol.


For simplicity let us first consider a complete cycle at the end
of which Alice receives from Bob a single-bit ``yes'' or ``no''
message. When Alice prepares her spin sample in the eigenstate $|
x+ \ra$, she  labels each spin as $i=1\, , ..N$.
From her $N$ weak  measurements of
${1\over \sqrt 2}(\sigma_x + \sigma_y)$, she  then  generates the
code by recording a string of real numbers corresponding to the
pointer readings $\{p_1,p_2,\, .. p_N \}$; the sample mean
$\bar{p} = {1 \over N}\sum_i p_i $ should yield the expectation
value $\la A \ra \simeq 1/\sqrt{2}$. Next the spins are given  to
Bob, carefully keeping track of the ordering. Now it's  his turn
to send the message. If Bob decides to send a ``yes'', he measures
 $\sigma_y$ on every single spin; otherwise he
measures $\sigma_z $  to signal a  ``no''. In either case, from
the resulting sequence of outcomes he generates the key: an
ordered  list of $N$ bits $\{k_1, k_2,\, ... \, k_n \}$ where,
say, $k_i=1$ and $0$ respectively correspond to the outcomes
``up'' and ``down'' for the the $i$-th spin. It is this key which
Bob  sends back to  Alice using an insecure channel. As the key is
effectively random by virtue of the fact that $\hat{y}$ and
$\hat{z}$  lie on the equal-probability plane, the uncorrelated
sequence of ``1'' and ``0'''s will be useless to the eavesdropper.

On the other hand, Alice, upon receiving the key, may go back to
her code and separate  each reading $p_i$ into either of two bins,
depending on whether  $k_i = 1$ or $0$. Finally,  she computes the
mean values of $p$ in each bins
\begin{equation}
 \bar{p}_1 = \frac{\sum_i p_i k_i}{\sum_i k_i}\, , \ \ \ \
\bar{p}_0 = \frac{\sum_i p_i (1 -k_i)}{\sum_i (1 - k_i)} \, .
\end{equation}
She takes these values as estimates of the ``true'' means within
errors of order $\Delta p/\sqrt{N/2}$. Now, if indeed Bob sent  a
``yes'', then she should see that  $\bar{p}_0 \simeq
A_w({y-})=0$, and  an ``eccentric'' weak value $\bar{p}_1 \simeq
A_w({y}+) = \sqrt{2}$. Instead, if Bob sent a ``no'', then
she will find no significant  deviation in either of the two means
$\bar{p}_1$ or $\bar{p}_0$;  for this case, the real part of
$A_w({z}\pm)$ coincides with the sample mean of
 $\la A \ra = 1/\sqrt{2}$. Thus, by distinguishing between
 a non-trivial and a trivial break-up of the sample mean,
Alice can decode Bob's single-bit message~\cite{extension}.

\begin{figure}
   \epsfxsize=3.2truein \epsfysize=6.5cm \centerline{\epsffile{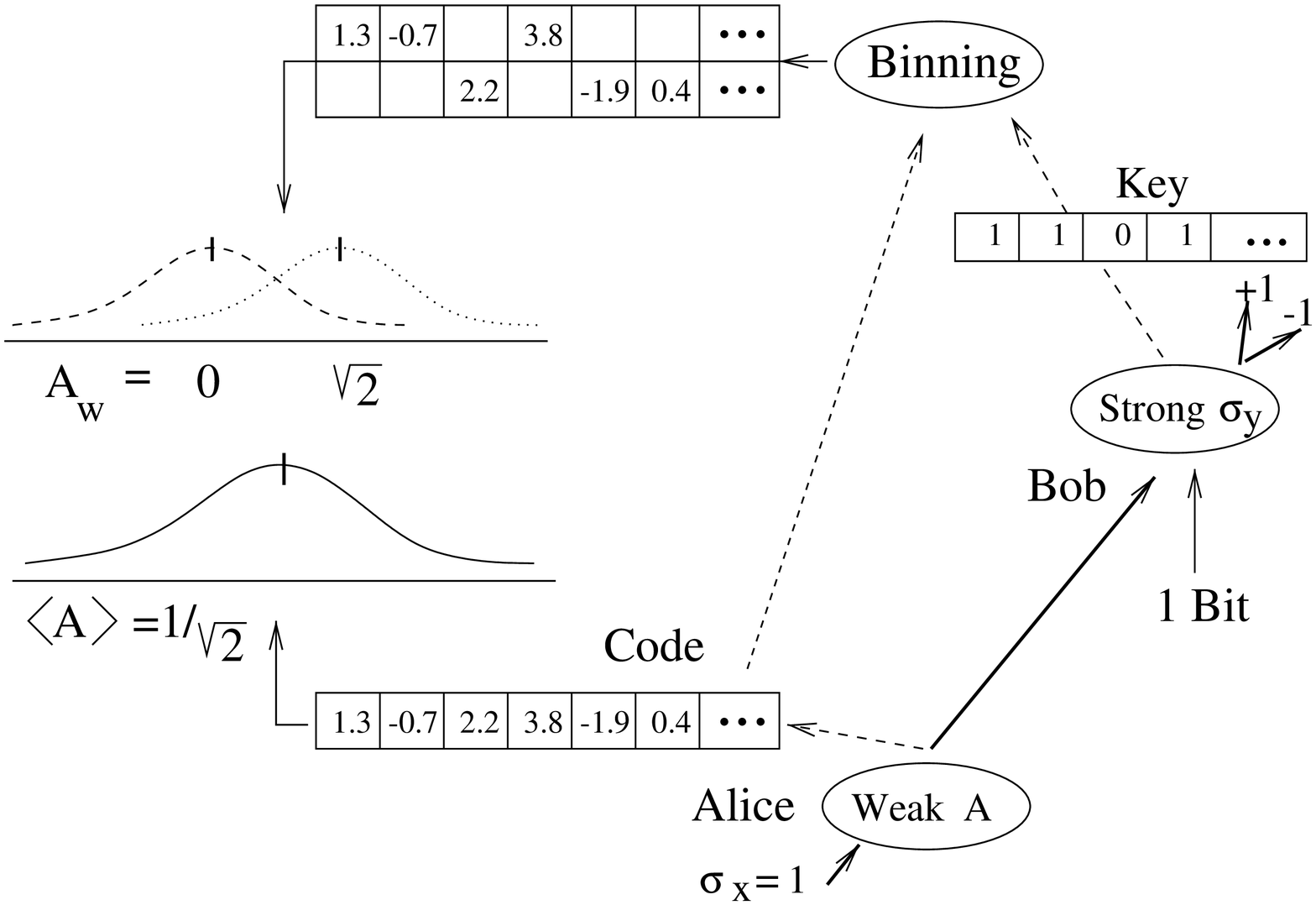}}
\caption[\, ]{ Schematic of a cycle from which Alice receives a
``yes''
 response from Bob.}
\label{thefig}
 \end{figure}

One may object that due to the small disturbance of the weak
measurements, the average numbers  of ``1'''s and ``0'''s in the
key  are no longer  identical for the two final measurements of
$\sigma_y$ and $\sigma_z$. Indeed, by eq. (\ref{freq}), there is
no change in the probabilities for the measurement along $z$, but
for the measurement along $y$ the relative change in the
probabilities is $\delta P({y}\pm) / P({y}\pm) = \pm 2 D $. The
difference may then be used by Eve to distinguish between the two
messages if the string is sufficiently long. This leads us to
compare the minimal sample sizes $N_E$ and $N_A$, respectively,
required by Eve and Alice to extract the message from their
available data. Alice needs to distinguish between the statistics
with means $\sqrt2$ and $0$. Hence she requires that $\sqrt2
\gtrsim {\Delta p/\sqrt{N_A/2}} $, which by eq. (\ref{D}) yields
$N_A \gtrsim {1/D}$. On the other hand, Eve must distinguish
between the relative frequencies of ``1'''s of $1/2$ for "no" and
$1/2\pm D$ for "yes", with an uncertainty in the frequencies
$\Delta f \simeq 1/2 \sqrt{N}$; for this she needs $N_E \gtrsim
{1/D^2 }$ trials. The ratio between the two optimal sample sizes
therefore scales like
$
{N_E \over N_A} \sim {1 \over D} \label{NfoverNw}
$
and consequently, for a sufficiently small $D$, an optimal value
$N$ may be chosen such that $N_E \gg N \geq N_A$ ensuring that the
key is safe.

While  this protocol is particularly interesting from the
conceptual point of view as we further discuss below,  it would
fail to meet usual cryptography security requirements  had Alice
sent the spins to Bob using a public channel. In such case, Eve could
have intercepted the message by performing weak measurements and
using the public key sent by Bob\cite{timerev}. The security
drawback is avoided in the second protocol, where by a similar set
of measurements, Alice sends a message to Bob.

As before, Alice starts with $N$ spins in the $|x+\ra$
state. To send a ``yes'' she next performs, on each spin
separately, a weak measurement of  $(\sigma_x+
\sigma_y)/\sqrt2$. To send a ``no'' she measures weakly
$(\sigma_x-\sigma_y)/\sqrt2$. In either case she records
the results as a code of $N$ real numbers, and sends to Bob
publicly {\em both} the code and the spins. To decode the message
Bob decides in a {\em random} manner to perform on each spin an
ordinary measurement, of either $\sigma_x$ or $\sigma_y$. The
$\sigma_x$ measurement is needed as a  ``security check'', while
the results of the $\sigma_y$ measurement will be used to decode
the message. If the spins  were not disturbed after
Alice's measurement, in $\sim N/2$ cases that Bob measures
$\sigma_x$ he gets $\sigma_x=1$ and only  $\sim ND\ll N/4$ times
$\sigma_x=-1$. On the other hand, he finds  $\sigma_y =\pm 1$ with
nearly equal probability. The latter are used to bin the code, into two
subsamples each with  $\sim N/4$ spins, corresponding to the
post-selection of $\sigma_y=1$  and $\sigma_y=-1$. Finally he
computes the mean value of each subsample. If Alice had measured
$(\sigma_x+\sigma_y)/\sqrt2$, Bob will find the weak values
$\sqrt2$ and $0$,  respectively. If instead she had measured
$(\sigma_x- \sigma_y)/\sqrt2$, he will find the same means but
in reverse order.

To discuss the question of security, notice that the weak
measurements rotate the spin to two slightly different spin
states. For a typical reading $p=1/\sqrt2\pm\Delta p$, the overlap
between the  two rotated states is of the order of $1-D$, as they
lie at angle of  $\sim\sqrt D$ from $x+$. Now, since Bob requires
$N\sim 1/D$ to  decode the message, the overlap between
 the two ``yes/no'' {\em collective}  $N$- spin  states becomes
 nearly orthogonal.  The crucial point is   that now there
 are two distinct ways to  distinguish between these states.
 The first, Bob's method of   post-selecting the spins separately,
 is not available to Eve; for instance, if she post-selects
 in the $\sigma_y$ direction,  Bob will notice that the mean of
$\sigma_x$ changed. An alternative is then for Eve to perform a
collective projection. However, in order to determine which
projector she needs to measure in the $2^N$-dimensional Hilbert
space, Eve {\em must} be in possession of the code {\em before}
she gets the spins. This information can therefore be concealed
from Eve if Alice sends the code {\em after} the spins have passed
by Eve, or if she sends the spins only one at a time. It can
further be shown that by separate spin measurements,
which rotate the spin by less than $\sqrt D$, Eve cannot gain
information. The security of the protocol, hence rests on two
ingredients: the non-commutativity of Bob's final measurements, as
in key distribution protocol \cite{BB}, and a temporal separation
between the spins and the code or between spins themselves. Such a
temporal delay was suggested by Goldenberg and Vaidman \cite{gv}.
However the latter proposal requires the use of a quantum storage
which is not needed in our case; Bob can complete his measurements
before receiving the code.

A simple quantum optical realization of the protocols
should be feasible following a previous proposal
~\cite{random,random2} for observing weak values through the
interaction of a coherent state in a high-Q resonant cavity
with a  pre- and post-selected Rydberg atom.
Here the photon occupation number
plays the role of the pointer variable.
Another realization may employ EPR-type correlations
in a photon-pair generated via
parametric down conversion process. In this case, one will perform
a weak-measurement on one photon and an ordinary measurement on
the other. Weak measurements of polarization have been
suggested~\cite{duck,knight} and carried out for classical
light~\cite{ritchie,story}. In the present case, one step further
is required, since the measurement must be performed on each
photon separately.

To conclude with, it is conceptually instructive to compare
the present suggestion to the  Bennett-Wiesner
communication scheme~\cite{bennett}: Alice prepares an EPR
pair and sends one of the particles to Bob. When Bob wishes
to send a 2-bit message to Alice, he performs a $\pi$
rotation of his spin  around the $x$,  $y$ or $z$ axis, or
does nothing. He then sends the spin back to Alice, who can
reveal his actions by measuring a joint observable.
 This scheme, as well as other  well known schemes
such as teleportation\cite{teleportation} or quantum key
distribution~\cite{BB,crypto}, ordinarily rely on
entanglement and its preservation ~\cite{purification}.
While in the present proposal the weak measurement can be
formulated in terms of entanglement between the measuring
device and the system, as in eq. (\ref{psipost}), the order
of events is here such that the entanglement is no longer
``there'' once the reading of the weak measurement has been
recorded.

How should we then understand the flow of information in the
present case? Bennett and Wiesner suggested after Schumacher, that
since in their scheme only one qbit is returned to Alice while two
bits of information are  transmitted,  `` {\it one bit of
information is  sent forward in time... while the other bit is
sent backwards in time to the EPR source, then forward in time
through the untreated particle}''. In our first example, because
the code is prepared before Bob sends his message, and because no
useful information can be extracted from the key, the message is
in some sense already ``in'' the  code. It seems therefore  that
the full one bit of information is sent backwards in time.  Yet,
it is only with the aid of the post-selected ``key'' that the
message is extracted from the quantum noise. Therefore, no
conflict arises between macroscopic causality and this apparently
retrocausal flow of information.

We thank Y. Aharonov, M. Byrd, A. Casher, T. Efron, M. Mims, Y.
Ne'eman, S. Nussinov, and L. Vaidman for helpful discussions, and
the Physics Department at the University of South
Carolina for their kind hospitality throughout the
preparation of the article. A. B. acknowledges the support
of Fondo Colciencias-BID and World Laboratory. B. R.  acknowledges
the support from grant 471/98 of the Israel Science Foundation,
established by the Israel Academy of Sciences and Humanities.

\noindent $\ast$  Also at {\em Centro Internacional de
F\'{\i}sica, Edificio Manuel Anc\'{\i}zar, Universidad Nacional,
Bogot\'{a},Colombia}.




\begin{thebibliography}{99}
\bibitem{directions}
A remarkable property exhibited by this sum rule is that the pair
of
 weak values
$\{ A_w(b),A_w(b-) \}$  will generally vary with the chosen final
direction $\hat{b}$, even when the probabilities $\{ P(b),
P(b-)\}$ remain
 unchanged under this variation.
\bibitem{extension}
An extension of the first protocol involves distinguishing $m$
different directions  on the equal probability
plane~\cite{directions}.
For large $m$  the optimal number
scales as $m^4$, meaning that the batch size grows exponentially
with the number of ``directional'' bits $\log_2 m$ per batch.
\bibitem{timerev}
 This would hold, in fact, even
if Alice had adopted a deception scheme in which the spins were
prepared at random in either the $|x +\ra$ or $|x- \ra$
directions. Alice would decode  the
message by binning into four instead of two categories.
However, Eve would still be able to decode the message
from the weak averages.
This is essentially a time-reversed version of a situation  in
which the initial states are distinguished by their corresponding
expectation values.
\end{thebibliography}
\end{document}